\documentstyle[12pt]{article}

\begin{document}
   \def\bea{\begin{eqnarray}}

   \def\eea{\end{eqnarray}}

\title{\bf {Effective Action for Self-Interacting Scalar Field in 3-dimensional Ball }}

\author{$^{1,2,3}$
M.R. Setare  \footnote{E-mail: rezakord@yahoo.com}, $^{1,2}$Kh.
Saaidi \footnote{ E-mail: ksaaidi@hotmail.com}
\\
 {$^{1}$Institute for Theoretical Physics and Mathematics,
Tehran, Iran}\\
{$^{2}$Department of Science, Physics group,
Kordestan University, Sanandeg, Iran}\\{$^{3}$ Department of
Physics, Sharif University of Technology, Tehran, Iran }}

 \maketitle

\begin{abstract}
In this paper we have considered the renormalized one-loop
effective action for massless self-interacting scalar field in the
3-dimensional ball. The scalar field satisfies Dirichlet boundary
condition on the ball. Using heat kernel expansion method we
calculate the divergent part of effective action, then by bag
model renormalization procedure we obtain the renormalized
one-loop effective action.
 \end{abstract}
\newpage

 \section{Introduction}
 The problem of calculating the determinant of a Laplacian-like
 operator $A$ on a manifold $M$ is very important in mathematics
 \cite{ray} and also in physics\cite{{Rem},{Eli},{bord}}. In the cases where $A$ has a
 discrete spectrum the determinant of $A$ is generally divergent.
 The zeta function regularization is an appropriate way for these
 calculations. The zeta function method is a particular useful
 tool for the determination of effective action, where one-loop
 effective action is given by $\frac{1}{2}\ln det A$. Using the
 relation between zeta function and heat-kernel for operator $A$,
 one can find the zeta function. Heat kernel coefficients play an important role in many
 areas of theoretical physics. In quantum field theory, heat kernel coefficients define
 the one-loop counter terms and quantum anomalies, as well as the large mass expansion of the
 effective action.\\
 In this paper we would like to consider a self-interacting massless
 scalar field which satisfies the Dirichlet boundary condition on a
 3-dimensional ball and calculate the relevant effective action.\\
 The study of a massless scalar field with quartic
 self-interaction is very important in different subject of
 physics, for example in the Winberg-Salam model of weak
 interaction, fermions masses generation, in solid state physics \cite{{stan},{bel}},
 inflationary models \cite {guth}, solitons \cite{{dash},{raj}}
 and Casimir effect \cite{{lang},{jos}}.\\
 The outline of the paper is as follows:
 In section 2 after a brief review of heat kernel expansion we
 calculate the heat kernel coefficients for operator $A=\Box+V''(
 \hat{\phi})$ in 3-dimensional ball. In section 3 we obtain the
 divergent part of effective action, and introduce the classical
 part of effective action, then using the bag model
 renormalization procedure \cite{{bord3},{bord4}} the renormaized
 one-loop effective action can be obtained. Section 4 is devoted
 to conclusions.

\section{Zeta function and Heat-Kernel coefficients }
Our aim is to derive the effective action of a self-interacting
massless scalar field on a 3-dimensional ball which is given by

 \begin{equation}
B^{3}=\{ x \in R^{3};|x| \leq R\},
  \end{equation}
  with boundary $S^{2}$, two dimensional sphere.
  The classical action is given by \cite{kirst}
\begin{equation}
S (\tilde{\phi}) =-\int_{B^{3}}[-1/2 \tilde{\phi}\Box
\tilde{\phi}+V(\tilde{\phi})] d^{3}x,
\end{equation}
where $\Box$, is the D'Alembert operator of the 3-dimensional ball
$B^{3}$ and $V(\tilde{\phi})$ is a potential of self interacting
scalar field. The above action has a minimum at
$\tilde{\phi}=\hat{\phi}$ which satisfies the classical equation
of motion
\begin{equation}
 -\Box \hat{\phi} +V'(\hat{\phi})=0.
\end{equation}
Quantum fluctuations $\phi=\tilde{\phi}-\hat{\phi}$ around the
classical background $\hat{\phi}$ satisfy the following equation
\begin{equation}
A \phi=(-\Box +V'' (\hat{\phi})) \phi=0.
\end{equation}
The effective action in the one-loop approximation is as follow
\begin{equation}
\Gamma^{(1)}=1/2 \ln det A/\mu^{2},
\end{equation}
where $\mu $ is an arbitary parameter with the dimensions of a
mass necessary from dimensional consideration. In the zeta
function regularization method the one-loop effective action (5)
is given by
\begin{equation}
\Gamma^{(1)}=-1/2 \zeta^{'}_{A/\mu^{2}}(0),
\end{equation}
where $\zeta _{A}(s)$ is the zeta function related to the operator
$A$ and is defined by
\begin{equation}
\zeta _{A}(s)=\sum_{j}\lambda_{j}^{-s}=\frac{1}{\Gamma
(s)}\int_{0}^{\infty} dt t^{s-1} tr K(x,x,t),
\end{equation}
where $K(x,x',t)$ satisfies the heat-kernel equation
\begin{equation}
(\frac{\partial}{\partial t}+ A)K(t,x,x')=0,
\end{equation}
with the initial condition
\begin{equation}
K(0,x,x')=g(x)^{-1/2}\delta(x,x').
\end{equation}
Here we impose Dirichlet boundary condition
\begin{equation}
K(t,x,x')|_{x\exists S^{2}}=0.
\end{equation}
The asymptotic expansion of the trace of the heat kernel is given
by \cite{avra}
\begin{equation}
Tr f K(t,x,x)=(4\pi t) ^{-3/2}tr (\int_{B^{3}}dx
\sqrt{g}\sum_{k=0}^{\infty}\frac{(-t) ^{k}}{k!}(f
a_{k})+\int_{S^{2}}d\theta \sqrt{\gamma}\sum_{k=0}^{\infty}
t^{\frac{k+1`}{2}}c_{\frac{k+1}{2}}(f))
\end{equation}
where $f(x)$ is an arbitary smooth function on the ball and
$\gamma= det \gamma_{ij}$, in which $\gamma_{ij}$ is the metric on
the boundary. The $a_{k}$ and $c_{\frac{k+1}{2}} $ are the heat
kernel coefficients. The heat kernel coefficients $a_{k}$ are
independent of the applied boundary conditions. The
$c_{\frac{k+1}{2}}$ coefficients depend on the boundary conditions
imposed. The several first and simplest $a_{k}$ coefficients are
given by \cite{avra}
\begin{equation}
a_{0}=1,
\end{equation}
\begin{equation}
a_{1}=Q-\frac{1}{6}R,
\end{equation}
 \begin{equation}
a_{2}=(Q-R/6) ^{2}-\frac{1}{3}\Box Q-\frac{1}{90}R_{\mu
\nu}R^{\mu \nu}+\frac{1}{90}R_{\mu \nu \alpha \beta}R^{\mu \nu
\alpha \beta}+\frac{1}{15}\Box R+\frac{1}{6}\tilde{R_{\mu
\nu}}\tilde{R^{\mu \nu}},
 \end{equation}
where $Q$ is a potential term, in our problem $Q$ is given by
 \begin{equation}
Q=V''( \hat{\phi}).
 \end{equation}
 As one can see the $a_{k}$ coefficients are functions of
 geometric quantities, $R_{\mu\nu\alpha\beta}$, $R_{\mu\nu}$ and
 $R$ are respectively, Rieman, Ricci and scalar curvature tensor.
 \begin{equation}
 \tilde{R}_{\mu\nu}=[ \nabla_{\mu}, \nabla_{\nu}],
 \end{equation}
where $\nabla_{\mu}$ is covariant derivative.
 Several first boundary coefficients in asymptotic expansion for
 Dirichlet boundary condition are as follow \cite{avra}
 \begin{equation}
c_{1/2}=-\frac{\sqrt{\pi}}{2},
 \end{equation}
 \begin{equation}
c_{1}=\frac{1}{3}K-\frac{1}{2}f^{(1)},
 \end{equation}
 \begin{equation}
 c_{3/2}=\frac{\sqrt{\pi}}{2}((\frac{-1}{6}\hat{R}-\frac{1}{4}R^{0}_{nn}+\frac{3}{32}K^{2}-
 \frac{1}{16}K_{ij}K^{ij}+Q)+\frac{5}{16}K
 f^{(1)}-\frac{1}{4}f^{(2)}),
 \end{equation}
where
\begin{equation}
f^{(1)}(z)=1/6+\frac{z^{2}}{6}(2+\frac{z^{2}}{2}-z(z^{2}+6)h(z)),
\end{equation}
\begin{equation}
f^{(2)}(z)=-1/6+\frac{z^{2}}{6}(-4+\frac{z^{2}}{2}-4z^{3}h(z)),
\end{equation}
\begin{equation}
h(z)=\int_{0}^{\infty}\exp -(x^{2}+2zx)
\end{equation}
Here $\hat{R}$ is the scalar curvature of the boundary, $K$ is
trace of extrinsic curvature tensor on the $S^{2}$,
\begin{equation}
K_{ij}=\nabla_{i}N_{j},
\end{equation}
where $N_{j}$ is outward unit normal vector.
 Now we rewrite
the Eq.(11) as follow
\begin{equation}
tr K(t)=(4\pi t) ^{-3/2}
\sum_{k=0,1/2,1,...}^{\infty}(\int_{B^{3}}dv a_{k}+\int_{S^{2}}ds
c_{k})\exp(-tV''(\phi))t^{k},
\end{equation}

The heat kernel coefficients of the Laplace operator on the
3-dimensional ball with Dirichlet boundary condition have been
calculated by Bordag et al \cite{bord}. In \cite{bord} the scalar
field is free, but in our problem, we have a self-interacting
scalar field with
\begin{equation}
V( \hat{\phi})=\frac{\lambda}{4!}\hat{\phi}^{4},
\end{equation}
therefore we must calculate the heat kernel coefficients for the
following operator
\begin{equation}
A=-\Box+V''(\hat{\phi})=-\Box+\frac{\lambda}{2}\hat{\phi}^{2}.
\end{equation}
Now we introduce the following heat kernel coefficients for
interacting case
\begin{equation}
B_{k}=(\int_{B^{3}}dv a_{k}+\int_{S^{2}}ds c_{k})(V''(
\hat{\phi})) ^{3/2-k}.
\end{equation}
 Using Eqs. (12-14), (17-19) and Eq.(27) one can find
\begin{equation}
B_{0}=\int_{B^{3}}dv
a_{0}V''^{3/2}=\int_{B^{3}}dv(\frac{\lambda}{2}\hat{\phi}^{2})^{3/2},
\end{equation}
\begin{equation}
B_{1/2}=\int_{S^{2}}ds
c_{1/2}V''=\frac{-\sqrt{\pi}}{2}\int_{S^{2}}(
\frac{\lambda}{2}\hat{\phi}^{2})ds,
\end{equation}
\begin{equation}
B_{1}=(\int_{B^{3}}dv a_{1}+\int_{S^{2}}ds
c_{1})V''^{1/2}=\int_{B^{3}}(
\frac{\lambda}{2}\hat{\phi}^{2})^{3/2}dv+\frac{2}{3R}\int_{S^{2}}
(\frac{\lambda}{2}\hat{\phi}^{2}) ^{1/2}ds
\end{equation}
\begin{equation}
B_{3/2}= \int_{S^{2}}ds c_{3/2}
=\frac{-\pi^{3/2}}{6}+\frac{\sqrt{\pi}}{2}\int_{s^{2}}\frac{\lambda}{2}\hat{\phi}^{2}ds,
\end{equation}
\bea B_{2}&=&(\int_{B^{3}}dv a_{2}+\int_{S^{2}}ds
c_{2})V''^{-1/2}=\int_{B^{3}}[\frac{\lambda^{2}}{4}\hat{\phi}^{4}\\
&-&
\frac{\lambda}{6}\Box\hat{\phi}^{2}](\frac{\lambda}{2}\hat{\phi}^{2})
^{-1/2}dv+\int_{S^{2}}\frac{4}{315 R^{3}}(
\frac{\lambda}{2}\hat{\phi}^{2})^{-1/2}ds,\nonumber
 \eea
 here $R$ is radius
of 3-ball.
 Now using Eq.(24),the zeta function, Eq.(7) has
the form
\begin{equation}
 \zeta_{A}(s)=\frac{1}{ (4\pi)^{3/2}
 \Gamma(s)}\sum_{k=0,1/2,1,...}^{\infty}\Gamma(s+k-3/2)(\int_{B^{3}}a_{k}dv+\int_{S^{2}}c_{k}ds)
 V''^{3/2-k-s}
 \end{equation}
 \section{The renormalize one-loop effective action}
Now we can find the one-loop effective action Eq.(6) for a space
with odd dimension. The unrenormalized one-loop effective action
is given by \cite{kirst},
\begin{equation}
 \Gamma^{(1)}=\frac{-1}{
(4\pi)^{3/2}
 }\sum_{k=0,1/2,1,...}^{\infty}\Gamma(k-3/2)(\int_{B^{3}}a_{k}dv+\int_{S^{2}}c_{k}ds)
 V''^{3/2-k}
 \end{equation}
 Using Eq.(27) we have
\bea
 \Gamma^{(1)}&=&\frac{-1}{ (4\pi) ^{3/2}}[ \Gamma(-3/2)\int_{B^{3}}(
\frac{\lambda}{2}\hat{\phi}^{2})^{3/2}dv-\Gamma(-1)\frac{\sqrt{\pi}}{2}\int_{S^{2}}(
\frac{\lambda}{2}\hat{\phi}^{2})ds+\Gamma(-1/2)[\int_{B^{3}}
(\frac{\lambda}{2}\hat{\phi}^{2})^{3/2}dv \nonumber \\
&+&\frac{2}{3R}\int_{S^{2}}
(\frac{\lambda}{2}\hat{\phi}^{2})^{1/2}ds]
+\Gamma(0)[\frac{-\pi^{3/2}}{6}+\frac{\sqrt{\pi}}{2}
\int_{s^{2}}\frac{\lambda}{2}\hat{\phi}^{2}ds] +
\Gamma(1/2)[\int_{B^{3}}[\frac{\lambda^{2}}{4}\hat{\phi}^{4} \\
&-&
\frac{\lambda}{6}\Box\hat{\phi}^{2}](\frac{\lambda}{2}\hat{\phi}^{2})
^{-1/2}dv+\int_{S^{2}}\frac{4}{315 R^{3}}(
\frac{\lambda}{2}\hat{\phi}^{2})^{-1/2}ds]+...].\nonumber
 \eea
 The
first five coefficients of heat kernel expansion contribute to
divergences of the one-loop effective action. Calling these first
five terms $\Gamma^{(1)}_{div} $, we have
\bea
 \Gamma^{(1)}_{div}&=&\frac{-1}{ (4\pi) ^{3/2}}[ \Gamma(-3/2)\int_{B^{3}}(
\frac{\lambda}{2}\hat{\phi}^{2})
^{3/2}dv-\Gamma(-1)\frac{\sqrt{\pi}}{2}\int_{S^{2}}(
(\frac{\lambda}{2}\hat{\phi}^{2})ds \nonumber \\
&+& \Gamma(-1/2)[\int_{B^{3}}(
\frac{\lambda}{2}\hat{\phi}^{2})^{3/2}dv+\frac{2}{3R}\int_{S^{2}}
\frac{\lambda}{2}\hat{\phi}^{2}) ^{1/2}ds]
+\Gamma(0)[\frac{-\pi^{3/2}}{6}+\frac{\sqrt{\pi}}{2}\int_{s^{2}}
\frac{\lambda}{2}\hat{\phi}^{2}ds]\nonumber \\
 &+&  \Gamma(1/2)[\int_{B^{3}}[\frac{\lambda^{2}}{4}\hat{\phi}^{4}-
\frac{\lambda}{6}\Box\hat{\phi}^{2}](\frac{\lambda}{2}\hat{\phi}^{2})
^{-1/2}dv+\int_{S^{2}}\frac{4}{315 R^{3}}(
\frac{\lambda}{2}\hat{\phi}^{2})^{-1/2}ds]].
 \eea
 At this stage we recall that $\Gamma^{(1)}$
is only one part of the total action. There is also a classical
part. We can try to absorb $ \Gamma^{(1)}_{div}$ into the
classical action. The classical action is
\begin{equation}
\Gamma_{class}=PV+\sigma S+FR+K+\frac{h}{R},
\end{equation}
where $V=\frac{4 \pi R^{3}}{3}$ and $S=4\pi R^{2}$ are,
respectively, volume of $B^{3}$ and surface of $S^{2}$, and $P$ is
pressure, $\sigma$ is surface tension  and $F,K,h$ do not have
special names. In order to obtain a well defined result for the
total one-loop effective action, we have to renormalize the
parameters of classical action according to below:
\begin{equation}
P \rightarrow P+\frac{1}{ (4\pi) ^{3/2}} \Gamma(-3/2)\int_{B^{3}}(
\frac{\lambda}{2}\hat{\phi}^{2}) ^{3/2}dv.
\end{equation}
\begin{equation}
\sigma \rightarrow \sigma- \frac{1}{(4\pi) ^{3/2}}\frac{1}{4\pi
R^{2}} \Gamma(-1)\frac{\sqrt{\pi}}{2}\int_{S^{2}}(
\frac{\lambda}{2}\hat{\phi}^{2})ds .
\end{equation}
\begin{equation}
F \rightarrow F+\frac{\Gamma(-1/2)}{ (4\pi)^{3/2}R}[\int_{B^{3}}(
\frac{\lambda}{2}\hat{\phi}^{2})^{3/2}dv+\frac{2}{3R}\int_{S^{2}}
(\frac{\lambda}{2}\hat{\phi}^{2}) ^{1/2}ds]
\end{equation}
\begin{equation}
K \rightarrow K+\frac{\Gamma(0)}{( 4\pi) ^{3/2}}
[\frac{-\pi^{3/2}}{6}+\frac{\sqrt{\pi}}{2}\int_{S^{2}}\frac{\lambda}{2}\hat{\phi}^{2}ds].
\end{equation}
\begin{equation}
h \rightarrow h+\frac{\Gamma(1/2)}{
(4\pi)^{3/2}}R[\int_{B^{3}}[\frac{\lambda^{2}}{4}\hat{\phi}^{4}-
\frac{\lambda}{6}\Box\hat{\phi}^{2}](\frac{\lambda}{2}\hat{\phi}^{2})
^{-1/2}dv+\int_{S^{2}}\frac{4}{315 R^{3}}(
\frac{\lambda}{2}\hat{\phi}^{2})^{-1/2}ds].
\end{equation}
Hence, the effect of the self-interacting scalar quantum field is
to change, or renormalize the parameter of classical part of
system. Using Eq.(37), the total action becomes
\begin{equation}
\Gamma^{tot}=\Gamma^{(1)}+\Gamma_{class},
\end{equation}
where the parameter of $\Gamma_{class}$ are given by Eqs.(38-42).
Once the terms named $\Gamma^{(1)}_{div}$ are removed from
$\Gamma^{(1)}$, the remainder is finite and will be called the
renormalized one-loop effective action
 \bea
\Gamma^{(1)}_{ren}&=&\Gamma^{(1)}-\Gamma^{(1)}_{div}=\frac{-1}{(4\pi)
^{3/2}}\sum_{k=5/2,3,...}\Gamma(k-3/2)(\int_{B^{3}}a_{k}dv+\int_{S^{2}}c_{k}ds)
V''^{3/2-k}\\
&=&\frac{-1}{(4\pi) ^{3/2}}\sum_{k=5/2,3,...}\Gamma(k-3/2)B_{k}
.\nonumber
\eea

\section{conclusion}
In this paper we have considered the renormalized one-loop
effective action for massless self-interacting scalar field in
3-dimensional ball. We assume the scalar field satisfies Dirichlet
boundary condition on the ball. Unlike to the main part of
previous studies on the scalar Casimir effect and one-loop
effective action here we adopt the boundary condition problem with
interacting quantum field. To obtain the divergent part of
one-loop effective action we calculate heat kernel coefficients
for operator $A=-\Box+V''( \hat{\phi})$. Previous result of heat
kernel coefficients for Laplace operator on the 3-dimensional ball
with Dirichlet boundary condition \cite{{bord},{bord2}} have been
deformed for interacting case in our problem. The new result are
given by Eqs.(28-32). The first five coefficients of heat kernel
expansion contribute to divergences of the one-loop effective
action. The renormalization procedure which is necessary to apply
in this situation is similar to that of the bag model
\cite{{bord3},{bord4}}. At this stage we introduce the classical
system and try to absorb divergent part into this classical
action, therefore we renormalize the parameters of classical
action. Extraction divergent part allows us to have the
renormalize one-loop effective action.

  \vspace{3mm}

{\large {\bf  Acknowledgement }}\\
We would like to thank Mr. A. Rezakhani for reading manuscript of
text.
 \vspace{1mm}
\small.

\end{document}